\documentclass[twocolumn,pre,nofootinbib]{revtex4}
\usepackage{amsmath,amssymb,graphicx}

\begin{document}
\title{
Metastate analysis of the ground states of two-dimensional Ising 
spin glasses
}
\author{A.K.~Hartmann}
\affiliation{Institut f{\"u}r Physik, Universit{\"a}t Oldenburg, 26111 Oldenburg,
Germany}
\author{A.P.~Young}
\affiliation{Department of Physics, University of California, Santa Cruz, California 95064}
\date{\today}
\begin{abstract}
Using an efficient polynomial-time ground state algorithm we investigate the
Ising spin glass state at zero temperature in two dimensions. For large sizes,
we show that the
spin state in a central region is independent of the interactions far away,
indicating a ``single-state" picture, presumably the droplet model.
Surprisingly, a single power law describes corrections to this result down to
the smallest sizes studied.
\end{abstract}

\maketitle

\section{Introduction}

Spin glasses  are prototypical disordered 
models \cite{binder:86,mezard:87,fischer1991,young:98} 
studied in statistical mechanics, with applications in various fields 
such as machine learning, neural networks, and 
optimization \cite{nishimori2001,phase-transitions2005,mezard:09,mooreC2011,kawashima2013}. 
Despite considerable effort, the nature of the spin glass state in three
dimensions below the
transition temperature $T_c$ remains uncertain. In equilibrium it is not possible to study
very large sizes numerically because relaxation is very slow in 
Monte Carlo \cite{newman:99}
simulations, and because a large number of samples have to averaged over.
Because we don't know how large are
corrections to the asymptotic scaling behavior, it is difficult to judge
whether results obtained in numerics show the asymptotic behavior or just a pre-asymptotic crossover.

The two main scenarios which have been proposed for the nature of the spin
glass state are the droplet 
picture \cite{fisher:86,fisher:88,mcmillan:84a,bray:86}
and the replica symmetry breaking (RSB) \cite{parisi:79,parisi:80,parisi:83} picture.
An important distinction between these two approaches is that the droplet
theory is a one-state picture, and RSB is a many-state picture.\footnote{We
work in zero magnetic field so states come in pairs, related by flipping
all the spins. Hence we should strictly use the terms ``one-pair" and
``many-pairs" rather than ``one-state" and ``many-states". However, here we 
prefer to 
use ``state" rather than ``pair". \label{fn1}} To explain
what this means we have to note that the state of the system can possibly
depend chaotically on system size. 
Because of this, to describe spin glasses we need the
``metastate" proposed by Newman and Stein (NS)
\cite{newman:97,newman:06} and by Aizenmann and
Wehr (AW) \cite{aizenman:90}. These two versions of the metastate are generally
thought to be equivalent but the AW metastate is more convenient for our
purposes so we will consider that here. For more information on the metastate
in spin glasses see e.g.~Ref.~\cite{read:14}.

The basic idea of the metastate is to measure 
how some part of the system, expressed
in terms of correlations, depends on changes in other parts of the system.
Ideas related to the metastate can be found in other works, were
distribution of window overlaps or probabilities of subsystem 
configuration changes induced by changes of the boundary conditions
were investigated 
 for spin glasses \cite{palassini:99a,manssen:15,middleton:99}, 
random-field systems \cite{middleton:02}
and others \cite{middleton:99}.
To construct the AW metastate, take a large system of linear
size $L$ and divide it into an inner region of size $M\  (< L)$ and an outer
region. Determine the state of the system, and record the spin correlations in a
small central region of size $K\ (\ll M)$ in the middle of the inner region. This is 
illustrated in Fig.~\ref{boxes} for $L = 8, M=4, K=2$. Then change the bonds in the outer
region only and recompute the correlations in the small central region.
Repeat this many times and, for $M, L \to \infty$, see if the central correlations are independent of
the outer bonds, which corresponds to a one-state picture, or whether they
change as the outer bonds are changed, which corresponds to a many-state
picture.

At least in zero magnetic field, numerics in three dimensions seems to favor a many-state
picture, see e.g.~\cite{billoire:17}, but supporters of the droplet picture suggest that there are large
corrections to scaling for the sizes which can be reached, and the observed
behavior is just a crossover.

The above is for three dimensions.  However, in two dimensions the situation is
different for two reasons. Firstly there is overwhelming evidence that the
transition only occurs at zero temperature \cite{mcmillan:84,bray:84,rieger:96,hartmann:01a}, and secondly there
are highly efficient polynomial-time algorithms for computing the ground state
\cite{bieche:80,hartmann:01,cecam2007,thomas2007,pardella2008} 
at least if there are periodic boundary conditions in no more than
one direction.  It is generally accepted, though not rigorously
proved except for a half-plane~\cite{arguin:10}, that the droplet picture
(a one-state picture) applies in two dimensions and more generally
when the transition is at $T=0$. Although there was some confusion in the
past due to an exponent relating energy to size being \textit{apparently} different for ``domain wall" and
``droplet" excitations, which would contradict the droplet theory, it was
later shown by one of us and Moore \cite{hartmann:02b,droplets_long2004} that this apparent
difference is due to
corrections to scaling being large for droplet excitations (though small for
domain walls), and for large enough sizes the exponents are the same. 

In this paper we use efficient ground state methods for a large range of sizes
to directly address the one-state versus many-state issue in two dimensions by investigating the
AW metastate at $T=0$. We find a one-state picture,
i.e.~consistent with the droplet theory, since, in the
limit of large system size, correlations at the center don't depend on the
bonds far away. Of particular interest is
whether there are large corrections to scaling which could prevent
this result being deduced from small sizes only. In fact there are not. The
extrapolation to infinite size follows a single power law down to very small
sizes. This is in contrast to certain other quantities for which large sizes
are needed to observe the true scaling 
behavior \cite{hartmann:02b,droplets_long2004}.

The plan of this paper is as follows. In Sec.~\ref{sec:method} we define the
model and the quantities we calculate. Our results are described in
Sec.~\ref{sec:results} and our conclusions summarized in
Sec.~\ref{sec:conclusions}.

\section{The Method}
\label{sec:method}

We use the standard Edwards-Anderson \cite{edwards:75} of the Ising spin glass
in zero field, for which the Hamiltonian is given by
\begin{equation}
\mathcal{H} = - \sum_{\langle i j} J_{ij} S_i S_j \, ,
\end{equation}
where the nearest-neighbor interactions $J_{ij}$ are independent Gaussian 
random variables with zero mean and
standard deviation $1$, and the $S_i$, which take values $\pm 1$, lie at the sites
of a square lattice of size $L \times L$. We work at $T=0$. Because the bond distribution
is continuous,
the ground state is unique,
apart from inversion of all the spins as discussed in footnote \ref{fn1}. 

We use a fast polynomial-time algorithm \cite{bieche:80,hartmann:01,cecam2007}  for which periodic
boundary conditions can be applied at most in one direction. In order to
preserve the symmetry of the square lattice we take free boundary conditions
in \textit{both} directions.

\begin{figure}
\includegraphics[width= 5cm]{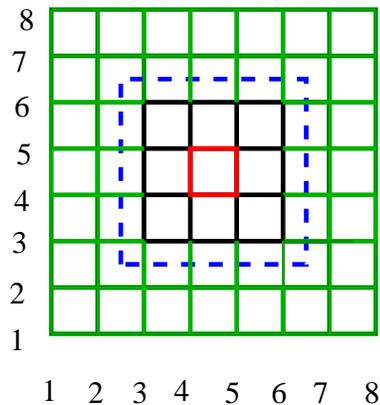}
\caption{(color online)
The setup for size $L=8$. The boundary between the
interior and exterior regions is indicated by the dashed blue line. For all sizes $L$,
the inner region is of size $M=L/2$, so $M=4$ here. 
Also, for all sizes we just measure the spin correlations in the central square (size $K=2$),
which is shown
in red. The rest of the interior region is shown in black. The exterior region,
whose bonds are changed after each run, is in green. Note that we take
free boundary conditions.
\label{boxes}
}
\end{figure}

The setup is shown in Fig.~\ref{boxes} for $L=8$ (we need $L$ to be a
multiple of $4$). The lattice is divided into ``inner" and ``outer" regions
separated by the (blue) dashed line in the figure. For all lattice sizes $L$
the inner region is of size $M = L/2$, so  
if we label a site by $(x, y)$ where $x$ and $y$ take values $1, 2, \cdots, L$,
then the values of $x$ and $y$ for sites in the inner region
have $x$ and $y$ in the range $L/4+1$ to $3L/4$. After determining the ground
state we record the four nearest-neighbor spin correlations for the spins on
the central square
(colored red in the figure) for which
the $x$ and $y$ values are
$L/2$ and $L/2+1$. 

Note that we fix the ratio $M/L$ to be $1/2$ as $L$ increases but 
the region where we compute the spin correlations is always just the central
square (so the size is $K=2$). Hence we consider the limit where $L/K$ and
$M/K$ become large with $L/M$ fixed.

\begin{figure}
\includegraphics[width=0.48\linewidth]{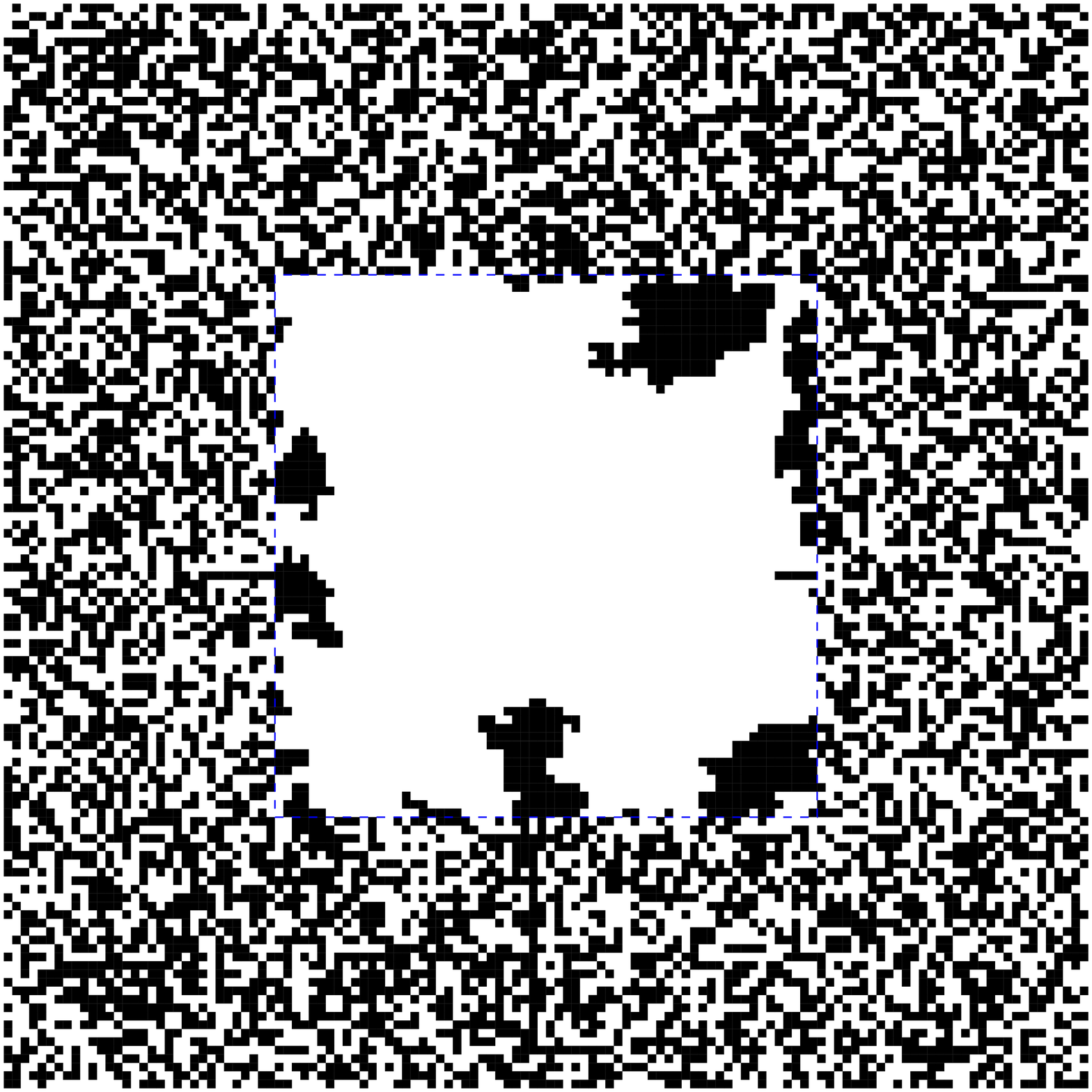}
\includegraphics[width=0.48\linewidth]{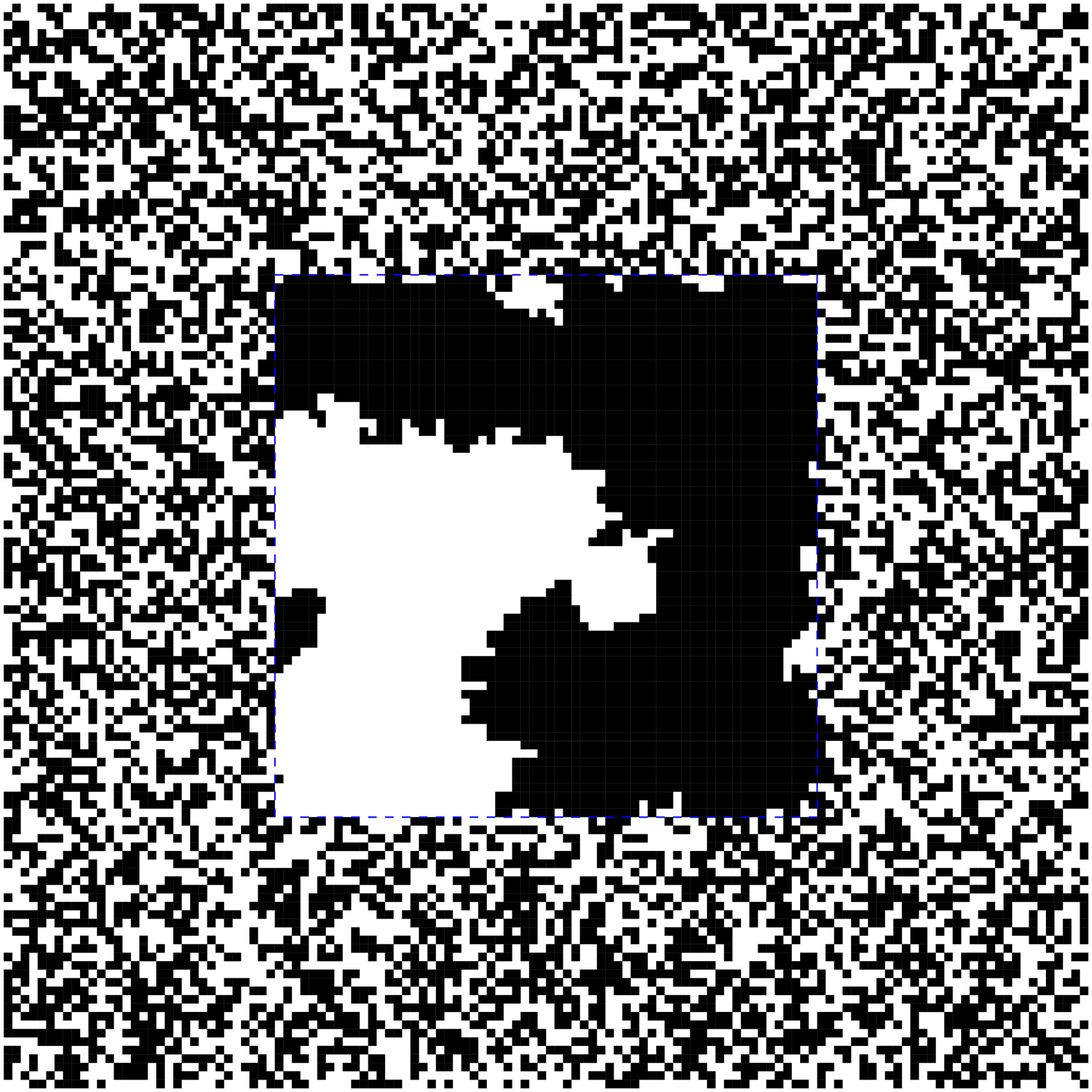}
\caption{
Comparison between pairs of ground states for $L=128$ 
which differ by changes of the bonds in the
outer region. Black squares indicate spins which a different. The left 
figure shows an example where few small clusters in the inner region are
flipped, while the right figure shows a flip of mainly one big cluster.
\label{fig:example:GS}
}
\end{figure}

Having recorded the spin orientations in the central square for a particular
choice of the bonds we then
change the bonds in the outer region only (shaded in
green in Fig.~\ref{boxes}) and recompute the ground state. 
This leads to strong changes of the configuration in the outer region,
basically half of the spins, and some changes in the inner region,
where the bonds have not changed. Two examples for the difference between
such pairs of ground states are shown in Fig.~\ref{fig:example:GS}. The
ground states in the inner region typically differ only by small spin clusters
or by one large cluster, respectively.

We repeat for many sets
of outer bonds but the same inner bonds and denote the corresponding
average by $[\ldots ]_\mathrm{out}$.
For $i$ and $j$ nearest-neighbor sites on the central square
we compute the average
\begin{equation}
C_{ij}^\mathrm{meta} = \Bigl|\, [S_i S_j ]_\mathrm{out} \Bigr| ,
\label{Cmeta_ij}
\end{equation}
which is known as a ``metastate" average. Here we take the modulus to get rid
of the random sign. Normally in spin glasses one takes the square, but there is
a reason discussed in connection with Fig.~\ref{Fig:cumul_dist_indiv}
why we prefer to take the modulus here. 
Almost identical results for the intercept $a$ and exponent $\lambda$ 
in Eq.~\eqref{fit} below are obtained if we use the square rather than the
modulus.
Note that the modulus is performed only after the average over the outer bonds
is done.
For each configuration of both inner and outer bonds $S_i
S_j = \pm 1$ since the ground state is unique apart from spin inversion.
Thus, if the state in the central region is completely 
independent of the outer
bonds one has $C_{ij}^\mathrm{meta} = 1$.
However, if the state in the central region 
depends on the outer bonds one has $C_{ij}^\mathrm{meta} < 1$.

Next we average over the inner bonds to get
\begin{equation}
C_\mathrm{meta, av} = [\, C_{ij}^\mathrm{meta}\, ]_\mathrm{in} =
[\, \Bigl| \, [ S_i S_j \, ]_\mathrm{out}\, \Bigr| \, ]_\mathrm{in} \,.
\label{meta-av}
\end{equation}
To get the best statistics we also average over
the four nearest neighbor pairs in the central square in
Fig.~\ref{boxes}.

According to a one-state picture 
\begin{equation}
C_\mathrm{meta, av} \to 1  \ \mathrm{for}\   L \to
\infty \quad \mathrm{(droplet\ theory)} ,
\end{equation}
while in a many-state picture, $C_\mathrm{meta, av}$
tends to a value less than $1$ in this limit.

\begin{figure}[!tbh]
\includegraphics[width= 8cm]{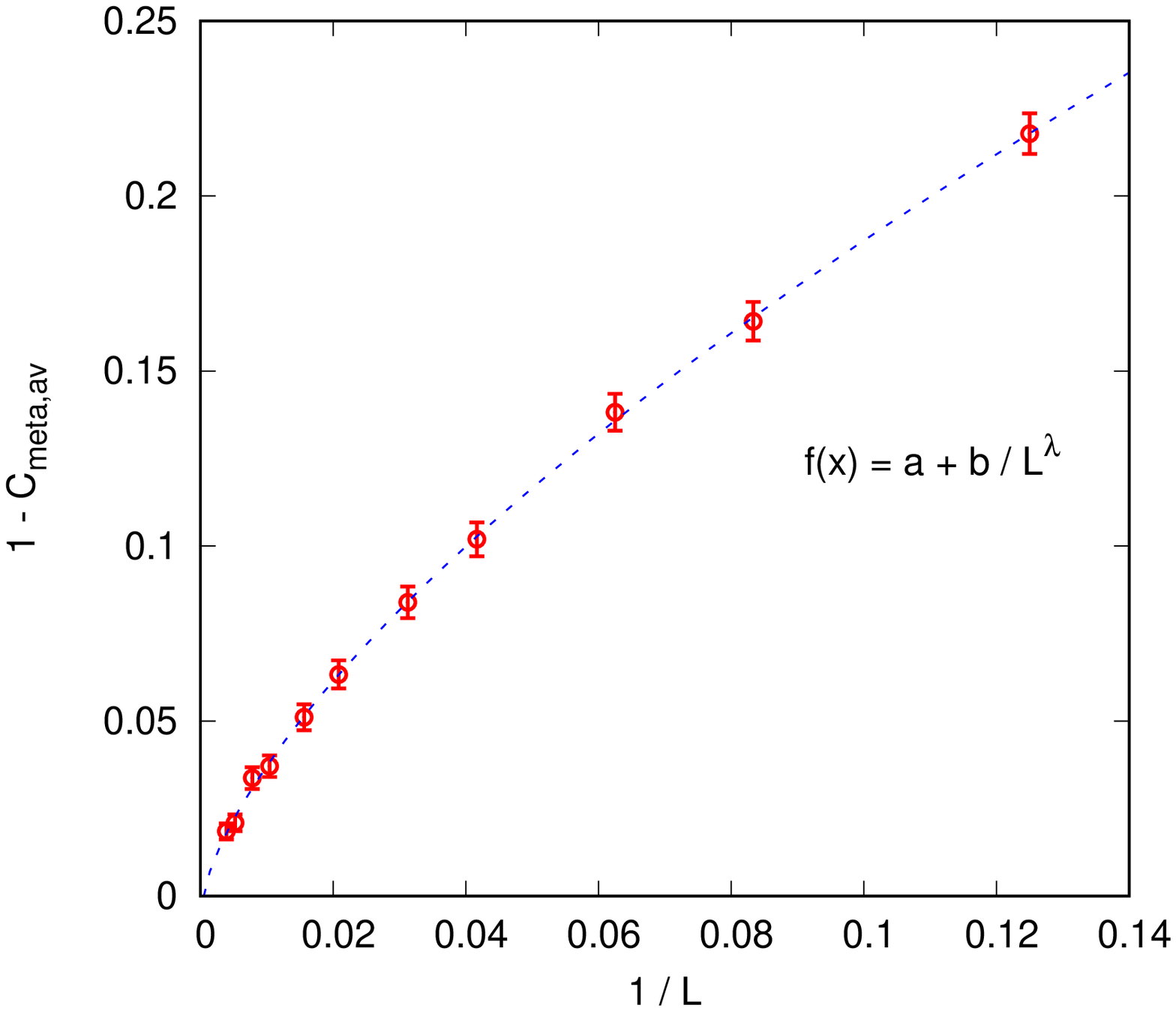}
\includegraphics[width= 8cm]{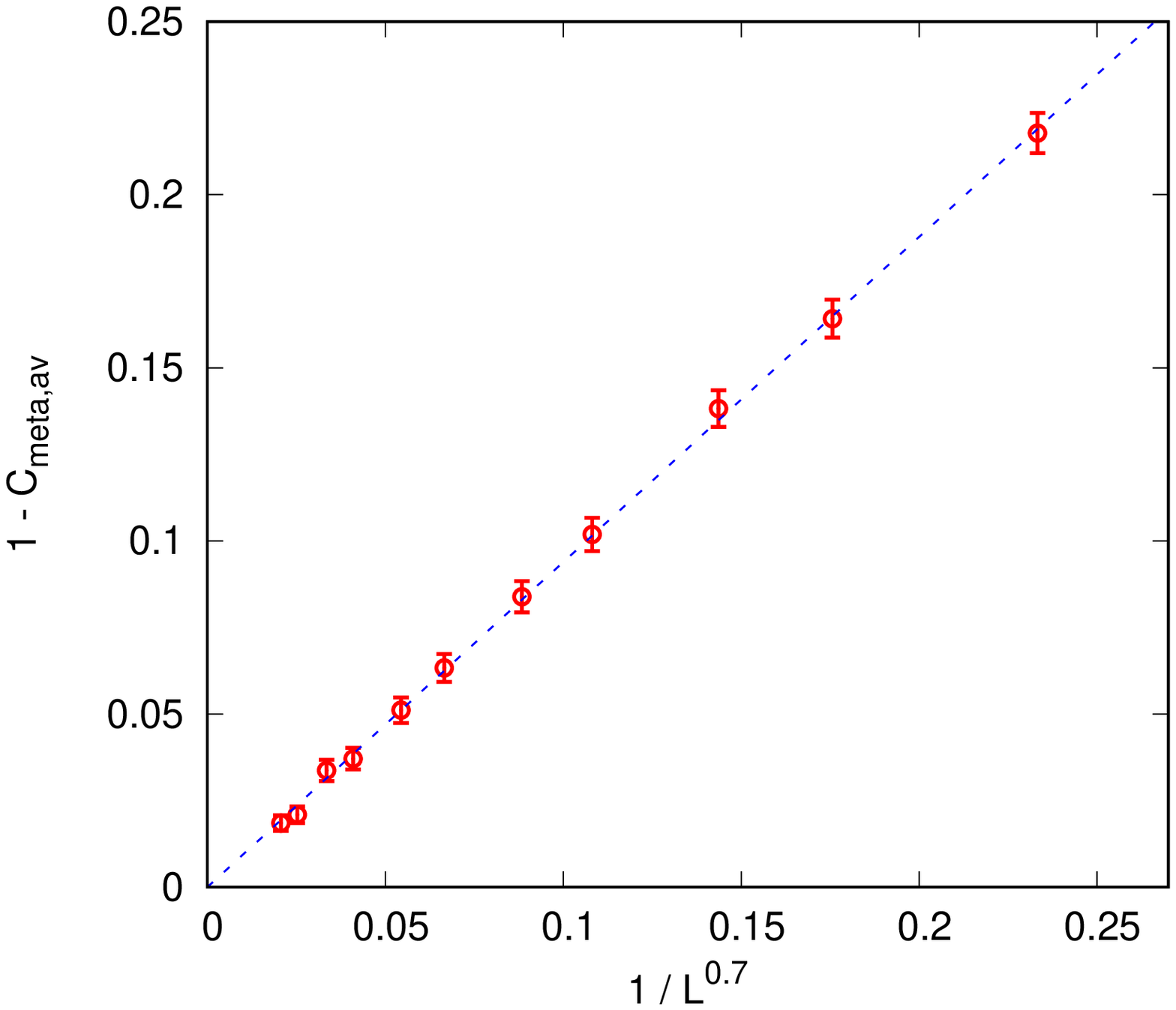}
\caption{
\label{Fig:meta-av} (color online)
The upper panel plots $1 - C_\mathrm{meta, av}$ against $1/L$ for different
sizes as well as a best fit according to Eq.~\eqref{fit}. The fit parameters
are $a = -0.004 \pm 0.005, \lambda = 0.66 \pm 0.05, b = 0.89 \pm 0.08$.
The lower hand panel plots the same data against $1/L^{0.7}$.
This power is chosen because the best fit with $a$ fixed to be $0$ gives
$c = 0.70 \pm 0.02$.  
}
\end{figure}

\section{Results}
\label{sec:results}

We have performed computations for sizes, $L=8$ to $L = 256$ using an efficient
polynomial time algorithm \cite{bieche:80,hartmann:01,cecam2007}
For each size we average over 100 choices of the outer bonds for a given
choice
of the inner bonds. This procedure is then repeated for 1000 values of the
inner bonds so altogether we do $10^5$ ground state computations for each
size.

\begin{figure}[!tbh]
\includegraphics[width= 9cm]{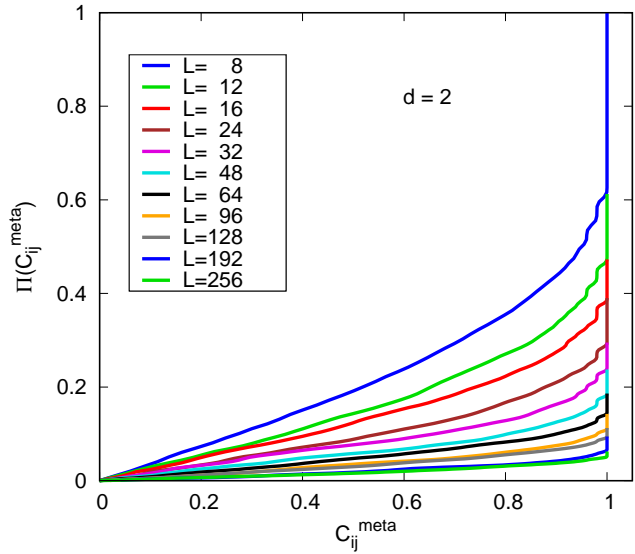}
\caption{(color online)
The cumulative probability distribution of $C^\mathrm{meta}_{ij}$ is plotted
for different sizes. If $P(C^\mathrm{meta}_{ij})$ is the probability
distribution of $C^\mathrm{meta}_{ij}$ averaged over both the inner bonds and
the four central nearest-neighbor pairs shown in Fig.~\ref{boxes}, then the
cumulative distribution $\Pi(y) $ is defined by $\Pi(y) = \int_0^y P(x)\, dx.$
The order of the lines in the figure is the same as in the legend. The 
distribution is actually discrete; the only values of
$C_{ij}^\mathrm{meta}$ which can occur are $0, 0.02, 0.04, \cdots, 1.00$ since
we take 100 sets of outer bonds for each set of inner bonds, and
$S_i S_j$ can only take the values $\pm 1$ at $T=0$. Note that the allowed
values of $C_{ij}^\mathrm{meta}$ are uniformly spaced. This is because we
defined $C_{ij}^\mathrm{meta}$ in Eq.~\eqref{Cmeta_ij} with a modulus. However,
if we had defined it with a square instead,
which would usually be more natural for spin glasses, the allowed values of
$C_{ij}^\mathrm{meta}$ would not be uniform.
\label{Fig:cumul_dist_indiv}
}
\end{figure}

The upper panel of Fig.~\ref{Fig:meta-av} plots $1 -
C_\mathrm{meta, av}$ against $1/L$.  To compute the error bars, we first
average over the four nearest neighbor pairs in the central square, to get a
single number for each sample. Since different samples have statistically
independent values 
for both the inner and outer bonds, the results for different samples are
statistically independent and so
error bars can be computed in the standard way from these results.
Also shown is a fit to the function
\begin{equation}
1 - C_\mathrm{meta, av} = a + {b \over L^\lambda} ,
\label{fit}
\end{equation}
which gives
\begin{align}
a &= -0.004 \pm 0.005 \\
b & = 0.89 \pm 0.08  \\
\lambda & = 0.66 \pm 0.05\, .
\end{align}
The quality  of this fit is $Q=0.99$,
very
close to $1$, which is a bit surprising. This may, partly, be a statistical coincidence,
and partly due to the values of
$C_{ij}^\mathrm{meta}$ not being Gaussian distributed (as can be inferred from
Fig.~\ref{Fig:cumul_dist_indiv}), so the \textit{true} fit probability can not be
obtained directly from the $\chi^2$ per degree
of freedom.

Our main result is that 
the extrapolated value $a$ is zero to within very small error bars which
provides
evidence that $C_\mathrm{meta, av} \to 1$ for $L\to\infty$
indicating a one-state picture. 

If we fix $a = 0$, a fit gives $\lambda = 
0.70\pm 0.02$ with the same quality $Q=0.99$ of the fit.
The lower
panel in Fig.~\ref{Fig:meta-av} plots $1 - C_\mathrm{meta, av}$ against
$1/L^{0.7}$. It is remarkable than an excellent straight-line fit is obtained
for all sizes from $L=8$ upwards. Hence 
single-state behavior could be deduced even for small sizes. However, one
would not be confident that this is the correct behavior without also
having the results
for large sizes. 

Note that the changes in the inner region are composed of clusters seperated
by domain walls, see Fig.~\ref{fig:example:GS}.
 Thus, the value of the exponent $\lambda$ is likely related 
\cite{palassini:99a,middleton:99,middleton:02} to the
fractal nature of the domain walls . The domain-wall lengths $l$ scale
 as $L^{d_\textrm{f}}$ where $d_\textrm{f}$ is the fractal dimension. 
Since the number of
affected bonds scales as $l$, the probability that a bond from the $O(L^2)$ 
bonds in the inner
region is affected should scale as $L^2/L^{d_\textrm{f}}=L^{2-d_\textrm{f}}$, i.e.,
$\lambda=2-d_\textrm{f}$. 
With estimates of $d_\textrm{f}=1.274(2)$ \cite{fract_dim_DW2007} and
$d_\textrm{f}=1.27319(9)$ \cite{Khoshbakht2018}
one has $\lambda \approx 0.73$ which compares well with 
our value $\lambda = 0.70\pm 0.02$ quoted above.

To get a more detailed picture, Fig.~\ref{Fig:cumul_dist_indiv} plots
$\Pi(C_{ij}^\mathrm{meta})$, the cumulative distribution of
$C_{ij}^\mathrm{meta}$, defined in Eq.~\eqref{Cmeta_ij}, averaged over the
different sets of inner bonds and the four central nearest-neighbor pairs 
in Fig.~\ref{boxes}. We see that
for a substantial
fraction of the choices
of the inner bonds one has $C_{ij}^\mathrm{meta}=1$. This 
fraction increases with increasing size and
apparently tends to 1 for $L \to \infty$. For these samples
there is strictly \textit{no}
change in the relative spin orientation of a central nearest-neighbor pair when the
outer bonds are changed. Note that the probability distribution of
$C_{ij}^\mathrm{meta}$, inferred from the cumulative distribution shown in 
Fig.~\ref{Fig:cumul_dist_indiv}, is very far from Gaussian, as discussed above in connection with with
$Q$-factor of the fits in Fig.~\ref{Fig:meta-av}.

\section{Conclusions}
\label{sec:conclusions}

We have confirmed the single-state picture for the $T=0$ spin glass state in $d=2$ by using
powerful numerical techniques which permit a study of large system sizes.
We have shown \textit{directly} that, at $T=0$ and for $d=2$,
the spin glass state in a region
only depends the bonds in the vicinity of that region and not on
the bonds far away.

An important result is that extrapolation to infinite system size is very smooth. A single inverse power
of $L$ describes the decay of
$1 - C_\mathrm{meta, av}$ to zero down to the smallest size studied $L=8$,
see Fig.~\ref{Fig:meta-av}.
Our estimates for this power, $\lambda$, are $0.66 \pm
0.05$ in an unconstrained fit, and $0.70 \pm 0.02$ if the extrapolated value
is fixed to zero.

The smooth trend with size shown in Fig.~\ref{Fig:meta-av} down to the 
smallest size $L=8$ is reminiscent of the domain wall energy. 
In early work, using sizes only up to $L = 12$, Bray and Moore\cite{bray:84}
found the stiffness exponent for the size dependence of domain wall
excitations
to be $\theta = -0.294 \pm 0.009$. Remarkably, this is very close to recent
results\cite{Khoshbakht2018} using efficient methods  
\cite{bieche:80,hartmann:01,cecam2007,thomas2007,pardella2008} which included
sizes between $L=8$ and $10,000$ and which found $\theta = -0.2793 \pm
0.0003$.
However the situation for the
energetics of droplet excitations is different, since for these
one needs quite large sizes to see the asymptotic behavior \cite{hartmann:02b}.
The reason for this difference is unclear. Unfortunately, it is also unclear
what sizes are needed in three dimensions to determine the nature of 
the spin glass
state on large length scales. Furthermore, the situation in three dimensions
is more complicated
because, in addition to the droplet and RSB pictures, there are additional
possibilities such as
like the ``trivial-non-trivial'' and ``chaotic-pair''
pictures \cite{newman2003}.

\acknowledgments
APY would like to thank the Humboldt Foundation for financial support and the
Institut f\"ur Physik, Universit{\"a}t Oldenburg for hospitality while
most of this work was carried out. 
We thank Nick Read, Dan Stein, and Mike Moore 
for a very helpful correspondence. 
  The simulations were performed at the
  the HPC cluster CARL, located at the University of Oldenburg
  (Germany) and
    funded by the DFG through its Major Research Instrumentation Program
    (INST 184/157-1 FUGG) and the Ministry of
    Science and Culture (MWK) of the
    Lower Saxony State.

\bibliography{refs}

\end{document}